\documentclass [aps,pra,twocolumn,superscriptaddress]{revtex4}
\usepackage{amsmath}
\usepackage{graphicx}

\begin{document}
\title{Robustness of the Rabi splitting under nonlocal corrections in plexcitonics}
\author{C. Tserkezis}
\affiliation{Department of Photonics Engineering, Technical University of Denmark, {\O}rsteds Plads 343, DK-2800 Kgs. Lyngby, Denmark}

\author{Martijn Wubs}
\affiliation{Department of Photonics Engineering, Technical University of Denmark, {\O}rsteds Plads 343, DK-2800 Kgs. Lyngby, Denmark}
\affiliation{Center for Nanostructured Graphene, Technical University of Denmark, {\O}rsteds Plads 343, DK-2800 Kgs. Lyngby, Denmark}

\author{N. Asger Mortensen}
\affiliation{Center for Nano Optics, University of Southern Denmark,\\ Campusvej 55, DK-5230 Odense M, Denmark}
\affiliation{Center for Nanostructured Graphene, Technical University of Denmark, {\O}rsteds Plads 343, DK-2800 Kgs. Lyngby, Denmark}

\begin{abstract}
We explore theoretically how nonlocal corrections in the description of the metal affect the strong coupling between excitons and plasmons in typical examples where nonlocal effects are anticipated to be strong, namely small metallic nanoparticles, thin metallic nanoshells or dimers with narrow separations, either coated with or encapsulating an excitonic layer. Through detailed simulations based on the generalised nonlocal optical response theory, which simultaneously accounts both for modal shifts due to screening and for surface-enhanced Landau damping, we show that, contrary to expectations, the influence of nonlocality is rather limited, as in most occasions the width of the Rabi splitting remains largely unaffected and the two hybrid modes are well distinguishable. We discuss how this behaviour can be understood in view of the popular coupled-harmonic-oscillator model, while we also provide analytic solutions based on Mie theory to describe the hybrid modes in the case of matryoshka-like single nanoparticles. Our analysis provides an answer to a so far open question, that of the influence of nonlocality on strong coupling, and is expected to facilitate the design and study of plexcitonic architectures with ultrafine geometrical details.
\end{abstract}
\maketitle

\section{Introduction}\label{Sec:intro}

The strong coupling of excitons in organic molecules and quantum dots with surface plasmons has been attracting increasing interest in recent years \cite{torma_rpp78,moerland_strongcoupling}. Both propagating surface plasmons in thin metallic films \cite{pockrand_jcp77,bellessa_prl93,dintinger_prb71,hakala_prl103,gomez_nl10,gonzalez_prl110,tumkur_oex24} or localised surface plasmons in metallic nanoparticles \cite{wiederrecht_nl4,sugawara_prl97,fofang_nl8,trugler_prb77,zengin_scirep3,zengin_prl114,melnikau_jpcl7} have been explored, not only from the point of view of theory, for which it constitutes a new class of light-matter interactions characterised by unique, hybrid optical modes \cite{lidzey_nat395,chang_prb76,delga_prl112,george_faraday178,tanyi_aom5}, but also because of its technological prospects in quantum optics and the emerging field of quantum plasmonics \cite{Bozhevolnyi_Springer2017,bozhevolnyi_nanophot} as well as the design of novel optical components \cite{hobson_admat14,dintinger_admat18,chang_natphys3,auffeves_pra79,schwartz_prl106}.

Originally, strong coupling effects were studied for atomic and solid-state systems inside optical or photonic cavities \cite{thompson_prl68,yoshie_nat432,aoki_nat443,faraon_natphys4}. $J$-aggregates of organic molecules are however increasingly more frequently preferred, as they are characterised by large dipole moments and narrow transition lines \cite{sugawara_prl97,lidzey_sci288}. At the same time, plasmonic architectures are excellent templates as cavities for the experimental realisation of hyprid exciton-photon systems, because they provide nanoscale confinement and small modal volumes \cite{schuller_natmat9}, thus permitting even single-molecule strong coupling at room temperature \cite{santhosh_natcom7,chikkaraddy_nat535}. The interaction of excitons with plasmons is usually traced through the Rabi splitting in optical spectra \cite{bellessa_prl93,sugawara_prl97}, but recent elaborate experiments showed that it is also possible to observe the coherent energy exchange between the optical states of the two components in real time through the corresponding Rabi oscillations\cite{vasa_natphot7,vasa_prl114,wang_afm26}. Tailoring the plasmonic environment is gradually departing from the regime of single nanoparticles \cite{fofang_nl8,trugler_prb77,waks_pra82,liu_prl118}, and dimers \cite{savasta_nn4,manjavacas_nl11,schlather_nl13,perez_nanotech25,roller_nl16,li_prl117,chen_nl17} or nanoparticle arrays \cite{wurtz_nl7,eizner_nl15,tsargorodska_nl16,todisco_nn10,li_jpcl7} are exploited to benefit from their stronger field confinement and  enhancement. A similar coupling has also been observed for the interaction of emitters with the phononic modes of SiC antennae \cite{esteban_njp16}, while engineering the electromagnetic vacuum through evanescent plasmon modes was recently proposed as a promising route towards increased light-matter interactions \cite{ren_prl117}.

As novel plasmonic architectures are explored in the quest for stronger, possibly even ultrastrong coupling \cite{geiser_prl108,balci_ol38,cacciola_nn8}, reducing characteristic lengths such as nanoparticle sizes and/or separations is a natural step that goes hand-in-hand with advances in nanofabrication. In such systems however, the excitation of a richness of hybrid optical modes characterised by large field intensities \cite{romero_oex14,raza_natcom6} is accompanied by the increased influence of nonclassical effects that require a description beyond classical electrodynamics \cite{zhu_natcom7}. Nonlocal screening due to the spatial dispersion of induced charges \cite{abajo_jpcc112,mcmahon_prl103,raza_prb84,raza_jpcm27}, enhanced Landau damping near the metal surface \cite{li_njp15,khurgin_faraday178,shahbazyan_prb94}, electron spill-out \cite{weick_prb74,scholl_nat483,monreal_njp15,toscano_natcom6,ciraci_prb93,christensen_prl118} and quantum tunnelling \cite{savage_nat491,scholl_nl13,yan_prl115} are different manifestations of the quantum-mechanical nature of plasmonic nanostructures in the few-nm regime, which need to be considered for an accurate evaluation of the emitter-plasmon coupling. For instance, nonlocal effects strongly affect spontaneous emission rates for quantum emitters in plasmonic environments \cite{filter_ol39}, and lead to reduced near-fields as compared to the common local-response approximation (LRA) \cite{raza_ol40}. Fluorescence of molecules in the vicinity of single metallic nanoparticles \cite{tserkezis_nscale8} or inside plasmonic cavities \cite{tserkezis_arxiv} is also affected by a combination of nonlocal induced-charge screening and surface-enhanced Landau damping. In the context of strong coupling, Marinica \emph{et al.} have reported quenching of the plexcitonic strength in dimers, attributed to quantum tunnelling \cite{marinica_nl13}. Even before entering the tunnelling regime, one anticipates that experimentally observed nonlocal modal shifts and broadening \cite{ciraci_sci337,raza_natcom6,shen_nl17} can manifest themselves as deviations in the dispersion diagram and changes in the width of the Rabi splitting in the optical spectra. Nevertheless, related studies are still missing and these expectations are yet to be confirmed.

In the recent review by T\"{o}rm\"{a} and Barnes \cite{torma_rpp78}, one section was devoted on how smaller dimensions of the plasmon-supporting structures become increasingly more important. In particular, it was anticipated that \emph{one consequence of nonlocal effects that is of great interest in the context of strong coupling is that of a reduction in the field enhancement that can be achieved when light is confined to truly nm dimensions}. Motivated by this discussion, and also by our recent studies of single emitters in plasmonic environments \cite{tserkezis_nscale8,tserkezis_arxiv}, we explore here the influence of nonlocality on plexcitonics through theoretical calculations based on the generalised nonlocal optical response (GNOR) theory \cite{mortensen_natcom5}. GNOR has proven particularly efficient in recent years in simultaneously describing both nonlocal screening and Landau damping through a relatively simple correction of the wave equation that is straightforward to implement to any plasmonic geometry model \cite{wubs_quantplas}. Here we take advantage of its flexibility to describe the optical response of typical plexcitonic architectures: metallic nanoshells either coated by an excitonic layer or containing it as nanoparticle core, dimers of such nanoshells and homogeneous nanosphere dimers encapsulated in an excitonic matrix. In all situations it is shown that, while nonlocal modal shifts introduce a detuning for fixed geometrical details, the width of the Rabi splitting remains in practice unaffected if one follows the more practical approach of modifying nanoparticle sizes and/or separations so as to tune the plasmon mode to the exciton energy. Broadening of the plasmon modes due to Landau damping is also shown to be weak enough, so as to secure that the two hybrid modes remain distinguishable, and linewidth-based criteria for strong coupling are satisfied. This somehow unexpected result is interpreted in view of a popular coupled-harmonic-oscillator (CHO) model and the strength of the coupling as related to the modal volume, while analytic solutions for the hybrid modes are obtained based on Mie theory. Our analysis relaxes concerns about nonlocal effects in the coupling of plasmons with excitons, and provides additional flexibility to the experimental realisation of novel plexcitonic architectures with sizes in the few-nm regime.

\section{Results and discussion}\label{Sec:results}

In order to facilitate a strong influence of nonlocality, we consider throughout this paper small nanoparticles with radii that do not exceed 20\,nm, and thin metallic shells of 1--2\,nm width. While such thin nanoshells are still experimentally challenging, successful steps towards this direction have been presented recently \cite{zhang_acsami8}. The choice of nanoshells over homogeneous metallic spheres serves a dual purpose. On the one hand, thin metallic layers ensure the reduced lengths required for a strong nonlocal optical response, even though the far-field footprint of nonlocality can often prove negligible \cite{tserkezis_jpcm20,raza_plasm8}. On the other hand, and more importantly, modifying the shell thickness provides the required plasmon tuning \cite{oldenburg_cpl288} to match the exciton energy and observe the anticrossing of the two hybrid modes in dispersion diagrams \cite{fofang_nl8,cacciola_nn8}. As discussed by T\"{o}rm\"{a} and Barnes \cite{torma_rpp78}, metals with low loss are beneficial for strong-coupling applications, because the plasmon modes should have linewidths comparable to those of the excitonic material. In our case, low intrinsic loss also ensures that the role of Landau damping will be better illustrated. For this reason, silver is adopted as the metallic material throughout the paper.

\begin{figure}[h]
\centerline{\includegraphics*[width=1\columnwidth]{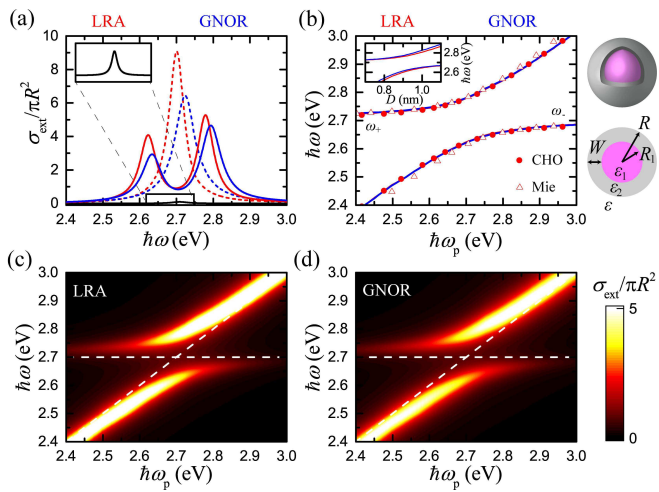}}
\caption{Strong coupling in the excitonic core-plasmonic shell nanoparticle shown in the schematics on the right. (a) Normalised extinction cross section ($\sigma_{\mathrm{ext}}$) spectra in the absence (dashed lines) or presence (solid lines) of the excitonic core characterised by a transition at energy $\hbar \omega_{\mathrm{exc}} =$ 2.7\,eV, for a silver nanoshell with radius $R =$ 5\,nm and shell thickness $W =$ 0.94\,nm (for which the plasmon energy $\hbar \omega_{\mathrm{p}}$ matches exactly $\hbar \omega_{\mathrm{exc}}$. Red and blue lines represent LRA and GNOR results, respectively. The black line (magnified in the inset) shows the corresponding spectrum for a homogeneous excitonic sphere with $R = $ 5\,nm. (b) Extinction peaks of the two hybrid modes ($\omega_{\pm}$) as a function of $\hbar \omega_{\mathrm{p}}$ within the LRA (red lines) and GNOR (blue lines) models. The plasmon energy is modified by changing the nanoshell thickness $W$, and is calculated through the extinction peaks in the absence of an excitonic core for each model. Dispersion diagrams of the two modes directly as a function of $W$ are shown in the inset. Red solid dots and open triangles denote the approximate solutions based on the CHO model and Mie-theory expansion, respectively, within LRA. Full extinction contour plots as a function of $\hbar \omega_{\mathrm{p}}$ calculated with LRA and GNOR are shown in (c) and (d) respectively, sharing a common colour scale. White dashed lines indicate the uncoupled exciton and plasmon energies.}\label{fig1}
\end{figure}

As a first example of a nonlocal plexcitonic system, we study a silver nanoshell of total radius $R =$ 5\,nm in air, encapsulating an organic-dye core of variable radius $R_{1}$, so that the metal thickness is $W = R - R_{1}$, as shown in the schematics of Fig.~\ref{fig1}. The dye is modelled as a homogeneous excitonic layer with a Drude--Lorentz dielectric function as described in Sec.~\ref{Sec:methods}, with its transition energy at $\hbar \omega_{\mathrm{exc}} =$ 2.7\,eV. The extinction ($\sigma_{\mathrm{ext}}$) spectrum for a homogeneous sphere of radius $R =$ 5\,nm described by such a permittivity is shown in Fig.~\ref{fig1}(a) by the black line (magnified in the inset), and is indeed characterised by a weak resonance at 2.7\,eV. With a red (blue) dashed line is depicted the corresponding LRA (GNOR) extinction spectrum for a hollow silver nanoshell (in the absence of the excitonic core) with $R = $ 5\,nm, and $W =$ 0.94\,nm, the thickness for which the symmetric particle-like plasmon mode of the nanoshell \cite{prodan_sci302} matches $\hbar \omega_{\mathrm{exc}}$ in the local description. As expected, when nonlocal effects are taken into account through GNOR, for the same geometrical parameters the plasmon mode experiences both a blueshift, of about 0.023\,eV, and additional broadening, so that the full width at half maximum of the resonance increases from 0.05\,eV to 0.067\,eV. Once the dye core is included, its excitonic mode interacts with the localised surface plasmon, and two hybrid modes are formed, as shown by solid red and blue lines in Fig.~\ref{fig1}(a), with a Rabi splitting of energy $\hbar \Omega_{\mathrm{R}} =$ 0.156\,eV. While both peaks experience a nonlocal blueshift, its strength is practically the same, so that $\hbar \Omega_{\mathrm{R}}$ remains unaffected. However, one should be cautious and not draw conclusions just from Fig.~\ref{fig1}(a), since the spectra have been calculated for a shell thickness for which the plasmon mode is detuned from $\hbar \omega_{\mathrm{exc}}$ in the nonlocal description.

In order to get a better visualisation of the strong coupling of Fig.~\ref{fig1}(a), we plot in Fig.~\ref{fig1}(b) the dispersion diagram of the two modes as a function of the plasmon energy $\hbar \omega_{\mathrm{p}}$, calculated by modifying the shell thickness and obtaining the resonance peak for hollow nanoshells. We then repeat the calculation in the presence of the dye, and obtain the frequencies of the hybrid modes, $\omega_{\pm}$, from the extinction spectra. Plotting the energies of the two hybrid modes as a function of $\hbar \omega_{\mathrm{p}}$ leads to dispersion lines that are practically indistinguishable for the LRA (red lines) and GNOR (blue lines) models. The same dispersion diagram is plotted in the inset of Figure~\ref{fig1}b as a function of shell thickness. It is straightforward to see that the GNOR dispersion lines are just horizontally shifted with respect to the corresponding LRA ones, as the nonlocal blueshift makes thicker shells necessary to achieve the same plasmon energy.

Together with the numerical results, in Fig.~\ref{fig1}(b) we also present the solutions obtained through two different analytic approaches: solid red circles represent the hybrid mode frequencies obtained from the CHO model, while open triangles correspond to analytic solutions based on Mie theory -- both are in excellent agreement with the full simulations. Due to its simplicity, the oscillator model is frequently adopted in literature to describe the modes in strongly-coupled plexcitonic geometries \cite{gomez_nl10,savasta_nn4,perez_nanotech25}. In the general case, the (complex) energies of the two hybrid modes are given by \cite{torma_rpp78,melnikau_jpcl7}
\begin{eqnarray}\label{Eq:CHOlossy}
&&\hbar \omega_{\pm} = \frac{\hbar \omega_{\mathrm{p}} + \hbar \omega_{\mathrm{exc}}}{2} - \mathrm{i} \frac{\hbar \gamma_{\mathrm{exc}}}{4} - \mathrm{i} \frac{\hbar \gamma_{\mathrm{p}}}{4} \nonumber\\
&&\pm \frac{1}{2} \sqrt{\left(\hbar \Omega_{\mathrm{R}} \right)^{2} + \left[\left(\hbar \omega_{\mathrm{p}} - \hbar \omega_{\mathrm{exc}}\right) + \mathrm{i} \left(\frac{\hbar \gamma_{\mathrm{exc}}}{2} - \frac{\hbar \gamma_{\mathrm{p}}}{2} \right)\right]^{2} }~,\nonumber\\
&&
\end{eqnarray}
where $\gamma_{\mathrm{p}}$ and $\gamma_{\mathrm{exc}}$ are the uncoupled plasmon and exciton damping rates, respectively. The frequency of the Rabi splitting is usually obtained by the simulated or experimental spectra at the crossing point, and it is directly related to the coupling strength $g$ through $\hbar \Omega_{\mathrm{R}} = 2 g$. 
When absorptive losses are low, and the plasmon and exciton linewidths are similar, Eq.~(\ref{Eq:CHOlossy}) simplifies to \cite{rudin_prb59}
\begin{equation}\label{Eq:CHO}
\hbar \omega_{\pm} = \frac{\hbar \omega_{\mathrm{p}} + \hbar \omega_{\mathrm{exc}}}{2} \pm \frac{1}{2 }\sqrt{\left(\hbar \Omega_{\mathrm{R}}\right)^{2} + \left( \hbar \omega_{\mathrm{p}} - \hbar \omega_{\mathrm{exc}} \right)^{2}}~.
\end{equation}
We note here that, since the width of the Rabi splitting is itself often obtained from the dispersion diagrams, instead of more elaborate, quantum-mechanical analyses based on the coupling strength, it is not that surprising that Eq.~(\ref{Eq:CHO}) reproduces the same data so well. On the other hand, for spherical particles one can obtain the position of the modes through the poles of the scattering matrix in the Mie solution \cite{tserkezis_arxiv}, an approach which is the nanoparticle equivalent to introducing the metal and excitonic dielectric functions into the surface plasmon polariton dispersion relation \cite{torma_rpp78}. This was done for example by Fofang \emph{et al.} \cite{fofang_nl8} in the simplified case where many of the (background) dielectric constants involved could be taken equal to unity. To generalise this, the dipolar Mie eigenfrequencies of a core-shell particle  can be obtained through \cite{tserkezis_arxiv}
\begin{equation}\label{Eq:ShellLSPs}
\varepsilon_{2} + 2 \varepsilon + \left(\frac{R_{1}}{R}\right)^{3} \frac{2 \left(\varepsilon_{1} - \varepsilon_{2}\right) \left(\varepsilon_{2} - \varepsilon\right)}{2 \varepsilon_{2} + \ell \varepsilon_{1}} = 0~,
\end{equation}
where $\varepsilon_{1}$ is the dielectric function of the core, $\varepsilon_{2}$ the corresponding one of the shell, and $\varepsilon$ describes the environment, as shown in the schematics of Fig.~\ref{fig1}. Using the Drude--Lorentz model discussed in the Methods section for $\varepsilon_{1}$, and a simple Drude model for $\varepsilon_{2}$ (with plasma frequency equal to 9\,eV and background dielectric constant equal to 3.65 to model silver), and disregarding absorptive losses in both materials as we are interested in the real resonance frequencies, leads to a relation in which $\omega$ is raised to the power of 6, and therefore six eigenfrequencies are obtained. Three of them are negative and have no physical meaning; the other three are the two hybrid plasmon-exciton modes, $\omega_{+}$ and $\omega_{-}$, shown by open triangles in Fig.~\ref{fig1}(b), and the antisymmetric, cavity-like plasmon mode of the nanoshell \cite{prodan_sci302}, which is at much higher frequencies and does not interact with the exciton.

Apart from the resonance frequency of the two hybrid modes and the width of the Rabi splitting, of particular interest is also the linewidth of the initial uncoupled modes, and that of the hybrid ones. In the description of mechanical harmonic oscillators, the condition for reaching the strong coupling regime is usually defined as $\Omega_{\mathrm{R}} > \omega_{\mathrm{mode}}$, where $\omega_{\mathrm{mode}}$ is the largest frequency between the two of the original uncoupled modes. In other contexts, this condition describes the ultrastrong coupling regime. In plasmonics these requirements are usually relaxed, and strong coupling is defined in a more pragmatic way, through \cite{torma_rpp78}
\begin{equation}\label{Eq:ConditionSC}
\hbar \Omega_{\mathrm{R}} > \sqrt{\frac{ \left(\hbar \gamma_{\mathrm{p}} \right)^{2}}{2} + \frac{\left(\hbar \gamma_{\mathrm{exc}} \right)^{2}}{2}}~.
\end{equation}
This is the condition required for the Rabi splitting to be observable in the spectra, and it indeed holds true in the example studied in Fig.~\ref{fig1}, as can be verified by introducing the linewidths of the uncoupled modes discussed above into Eq~(\ref{Eq:ConditionSC}). To better illustrate the broadening and damping of the modes predicted by nonlocal theories, in Figs.~\ref{fig1}(c)-(d) we show contour plots of extinction versus energy, obtained with the LRA and GNOR model, respectively. For all shell thicknesses (and therefore the corresponding plasmon energies) studied here, the two hybrid modes are clearly distinguishable, and the relaxed strong-coupling condition of Eq.~(\ref{Eq:ConditionSC}) holds. We have repeated the same analysis for several nanoparticle sizes and radius--thickness combinations, even reproducing results in the ultra-strong coupling regime \cite{cacciola_nn8}, without observing qualitative or quantitative differences from the above discussion.

\begin{figure}[h]
\centerline{\includegraphics*[width=1\columnwidth]{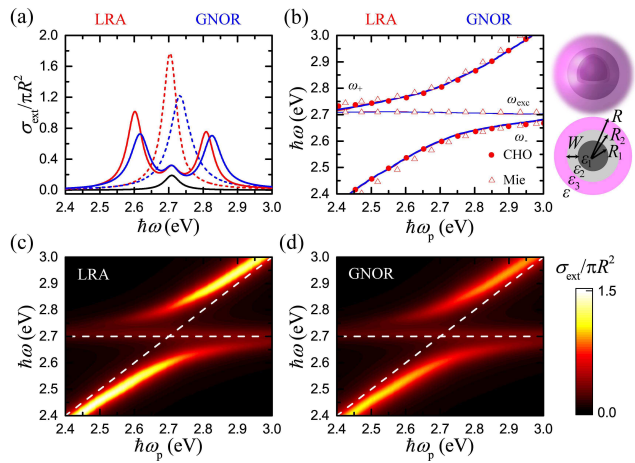}}
\caption{(a) Dipolar particle-like plasmon mode for a nanoshell with $R_{1} =$ 5\,nm, $W =$ 1.22\,nm (see schematics on the right) calculated within LRA (red lines) and GNOR (blue lines) in the absence (dashed lines) or presence (solid lines) of a concentric homogeneous excitonic layer with $R =$ 10\,nm and $\hbar \omega_{\mathrm{exc}} =$ 2.7\,eV. The black solid line depicts the excitonic resonance of the corresponding homogeneous excitonic sphere ($R =$ 10\,nm). (b) Energies of the hybrid modes $\omega_{\pm}$ as a function of $\hbar \omega_{\mathrm{p}}$, within LRA (red lines) and GNOR (blue lines). Red solid dots and open triangles represent the approximate solutions of the CHO model and the Mie expansions, respectively. The thin nearly horizontal line at 2.7\,eV is the uncoupled excitonic resonance supported by the outer surface of the particle. Extinction contour plots, and the corresponding uncoupled plasmon and exciton energies (white dashed lines), are shown in (c) and (d) for the LRA and GNOR model respectively. The two contours share a common colour scale.}\label{fig2}
\end{figure}

As a second example, it is natural to explore the inverse topology, where the excitonic layer grows around a metallic nanoshell, a geometry which is easier to achieve experimentally and has been shown to be preferable for obtaining large Rabi splittings and even entering the ultrastrong coupling regime \cite{cacciola_nn8}. In Fig.~\ref{fig2} we replace the excitonic core of the nanoshell of Fig.~\ref{fig1} with a SiO$_{2}$ one (described by a dielectric constant $\varepsilon_{1} = 2.13$). We assume a constant nanoshell radius $R_{2} =$ 5\,nm, covered by an external excitonic layer of thickness $R - R_{2} =$\,5\,nm and exciton energy 2.7\,eV as previously, and modify the thickness $W$ of the nanoshell, as shown in the schematics of Fig.~\ref{fig2}. The extinction spectra of the uncoupled components are shown in Fig.~\ref{fig2}(a), with a black line for the spectrum of a homogeneous excitonic sphere of $R = $ 10\,nm, and with red (blue) dashed lines for the LRA (GNOR) spectra of a silver nanoshell with $W =$ 1.22\,nm. A thicker shell than in Fig.~\ref{fig1}(a) is required to bring the plasmon resonance at 2.7\,eV, since a higher-index dielectric is now used as the core material. The corresponding coupled spectra are shown by solid lines, and are characterised by a Rabi splitting $\hbar \Omega_{\mathrm{R}} =$ 0.206\,eV. In addition to the two hybrid modes, a third peak of much lower intensity is visible at 2.7\,eV, which does not shift at all under nonlocal corrections. This is the surface exciton polariton mode excited at the outer surface of the dye layer, which indeed has no reason to be affected by the plasmon modal shifts \cite{philpott_mclc50}. It is always excited for this kind of nanoparticle topology, but it is usually masked by the much stronger peaks of the hybrid modes due to the large size of the plasmonic particles. The small nanoshell dimensions explored here ensure that it is clearly discernible in the extinction spectra. Such modes were recently explored as substitutes for plasmons in metals for applications in nanophotonics \cite{gentile_jopt19}.

In Fig.~\ref{fig2}(b) we present the corresponding dispersion diagram, as a function of the plasmon energy, similarly to Fig.~\ref{fig1}(b). Given our previous analysis in relation to Fig.~\ref{fig1}(b), and the spectra presented in Fig.~\ref{fig2}(a), it is not surprising that the dispersion lines for the LRA and GNOR models are again indistinguishable. The CHO model captures excellently the two hybrid modes, as shown by the red solid dots, but it lacks a description of the uncoupled excitonic mode. This weak resonance is excellently reproduced however by the corresponding Mie-based analytic solutions. Since in this case we study a three-layer nanoparticle, one should in principle obtain the corresponding scattering matrix \cite{Bohren_Wiley1983,stefanou_spie6989} and use appropriate asymptotic expressions \cite{Arfken_Academic2000} to obtain the resonances from its poles. A simpler approximation, which we successfully adopt here, is to separate the nanoparticle into two components: a metallic nanoshell embedded in an infinite excitonic environment described by a dielectric function $\varepsilon_{3}$, for which the hybrid modes $\omega_{\pm}$ can be obtained from Eq.~(\ref{Eq:ShellLSPs}) (replacing $\varepsilon$ with $\varepsilon_{3}$), and an excitonic sphere of $\varepsilon_{3}$ in an infinite environment described by $\varepsilon$, for which the resonance condition is \cite{tserkezis_arxiv}
\begin{equation}\label{Eq:SphereLSPs}
\varepsilon_{3} + 2\varepsilon = 0~.
\end{equation}
As can be seen by the open triangles in Fig.~\ref{fig2}(b), due to the small sizes involved, this approach works extremely well to accurately describe not only the hybrid modes $\omega_{\pm}$, but the uncoupled excitonic mode as well. Finally, in Figs.~\ref{fig2}(c)-(d) it is shown through full extinction contours that the additional nonlocal broadening of the modes is not significant in this case either. Considering the initial, uncoupled linewidths of the plasmon resonances shown in Fig.~\ref{fig2}(a) (0.049\,eV for LRA and 0.072\,eV for GNOR), Eq.~(\ref{Eq:ConditionSC}) again implies that strong coupling is still achievable under nonlocal corrections.

\begin{figure}[h]
\centerline{\includegraphics*[width=1\columnwidth]{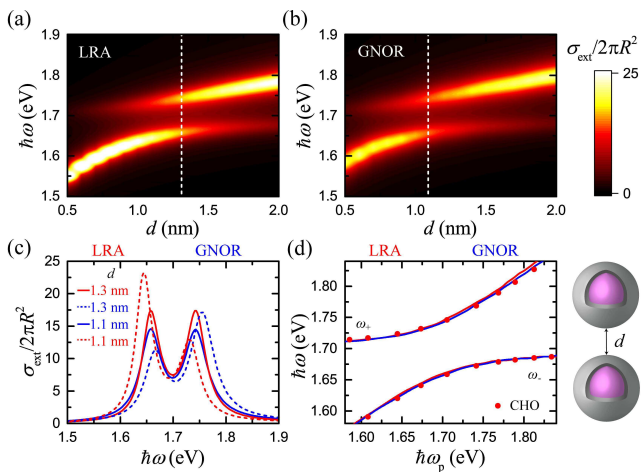}}
\caption{Extinction contour plots as a function of dimer gap width $d$ for the exciton core--silver shell dimers shown in the schematics on the right ($R =$ 20\,nm, $W =$ 2\,nm), within the LRA (a) and GNOR (b) models. Vertical dashed lines indicate the gap for which the bonding dimer plasmon matches the energy of the exciton ($\hbar \omega_{\mathrm{exc}} =$ 1.7\,eV). (c) Extinction spectra obtained with the LRA (red lines) and GNOR (blue lines) models, in the presence of an excitonic core, for the gap for which the particular model predicts a plasmon resonance at 1.7\,eV (solid lines) and for the $d$ for which the other model is tuned (dashed lines). (d) Dispersion diagram of the two hybrid modes as a function of $\hbar \omega_{\mathrm{p}}$, which is obtained for each model by modifying $d$ in the absence of an excitonic nanoparticle core. Red dots indicate the solutions of the CHO model in the LRA case.}\label{fig3}
\end{figure}

In search for a geometry where the footprint of nonlocality might be stronger, we depart from the description of single nanoparticles, and study in Figure~\ref{fig3} an exciton core--silver shell dimer. The particles have a total radius $R =$ 20\,nm and a metallic shell of thickness $D =$ 2\,nm, and are separated by a variable gap of width $d$. Both the thin shell thickness and the narrow dimer gap are expected to enhance nonlocal effects in this case. Instead of tuning the plasmon through the shell thickness, as in Figs~\ref{fig1} and \ref{fig2}, here we change the width of the gap. Plasmon hybridisation \cite{prodan_sci302} causes the bonding dimer plasmon modes to redshift as the gap shrinks \cite{romero_oex14}, allowing for efficient tuning of the optical response. Extinction contours obtained with the LRA and GNOR models as a function of $d$ are shown in Figs.~\ref{fig3}(a) and (b), respectively, for an excitonic core with $\hbar \omega_{\mathrm{exc}} =$ 1.7\,eV. The vertical dashed lines denote the gap for which the bonding dimer plasmon mode in the absence of a dye core is at 1.7\,eV, and indicate a 0.22\,nm difference between the two models. In Fig.~\ref{fig3}(c) extinction spectra for each model are presented both for this gap width within the corresponding model (solid lines), and also for the gap which tunes the plasmon at 1.7\,eV within the other model (dashed lines). Examining the tuned situation for each model, it is evident that no change in the width of the Rabi splitting and no significant broadening can be observed for this geometry either. This is more clear in the dispersion diagram of Fig.~\ref{fig3}(d), where the energy of the hybrid modes is plotted versus the plasmon energy in each model, leading once more to almost identical lines, also in good agreement with the CHO model.

\begin{figure}[h]
\centerline{\includegraphics*[width=1\columnwidth]{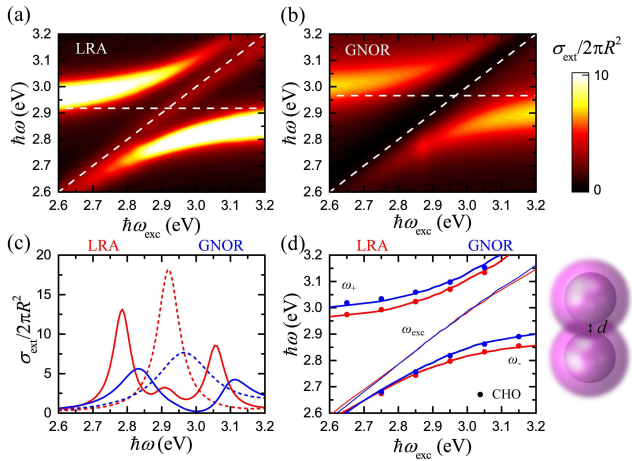}}
\caption{Extinction contour plots as a function of excitonic energy $\hbar \omega_{\mathrm{exc}}$ for a silver nanosphere ($R =$ 15\,nm) dimer ($d =$ 1\,nm) encapsulated in an excitonic matrix, as shown in the schematics on the right, within the LRA (a) and GNOR (b) models. The white dashed lines indicate the uncoupled plasmon and exciton energies. (c) Extinction spectra obtained within LRA (red lines) and GNOR (blue lines), in the absence (dashed lines) or presence (solid lines) of the excitonic matrix. (d) Dispersion diagram of the two hybrid modes as a function of $\hbar \omega_{\mathrm{exc}}$. The thin diagonal lines indicate the uncoupled excitonic resonance supported by the outer surface of the excitonic layer. Red and blue solid dots indicate the solutions of the CHO model in the LRA and GNOR case, respectively.}\label{fig4}
\end{figure}

Finally, we turn again to the inverse topology, of a dimer enclosed in an excitonic matrix, as shown schematically on the right-hand side of Fig.~\ref{fig4}. To avoid repetitions and make things more interesting, instead of nanoshells we consider homogeneous silver spheres of radius $R =$ 15\,nm separated by a gap of $d =$1\,nm, covered with an excitonic layer made of two spheres of radius 20\,nm, overlapping around the gap as they are concentric with the corresponding silver particles. In the absence of the excitonic layer the plasmon resonance is found to be at 2.92\,eV within LRA, and at 2.97\,eV within GNOR. Instead of tuning the dimer plasmon through some geometrical parameter, we keep in this case the geometry fixed and modify $\hbar \omega_{\mathrm{exc}}$ instead, an approach often adopted in theoretical calculations due to its convenience \cite{perez_nanotech25}. Extinction contour maps obtained with LRA and GNOR are shown in Figs.~\ref{fig4}(a) and (b) respectively, together with the uncoupled excitonic and plasmonic lines (white dashed lines). The corresponding extinction spectra in the presence (solid lines) or absence (dashed lines) of the excitonic layer are shown in Fig.~\ref{fig4}(c). One again observes that the width of the Rabi splitting appears to be nearly unaffected, but the hybrid modes are significantly broadened. Furthermore, the uncoupled excitonic peak, which is strong within LRA, nearly vanishes within GNOR, and can only be traced through the corresponding scattering spectra. Nevertheless, this strong damping, originating mainly from the presence of a much larger amount of metallic material (homogeneous spheres instead of shells), is still not enough to bring the system outside the strong coupling regime as defined by Eq.~(\ref{Eq:ConditionSC}). In addition, careful examination of the dispersion diagrams in Fig.~\ref{fig4}(d) shows that the Rabi splitting is in fact narrower by 0.012\,eV in the case of GNOR, finally identifying a small impact of nonlocality on the plexcitonic response.

In all the examples explored above, we focused on the dominant, dipolar plasmon modes, which are more relevant from an experimental point of view. While we have identified situations in which our conclusions hold for higher-order modes as well, the coupling of excitons with e.g. quadrupolar single-particle or bonding-dimer modes, and how it is affected by nonlocal effects, relies on several parameters, and a more systematic study is required. Efficient excitation of higher-order modes in single nanoparticles calls for larger particle sizes, for which nonlocal effects tend to become negligible \cite{raza_natcom6}. On the other hand, for very thin metallic shells, nonlocality can wash out higher-order modes \cite{tserkezis_nscale8}, so that a classically predicted Rabi splitting could disappear completely. Similarly, the efficient excitation of quadrupolar bonding-dimer modes might require to enter the truly sub-nm regime \cite{romero_oex14}, where quantum tunnelling prevails, restricting nonlocality to a secondary role. Further theoretical work in the future should shed more light on such issues.

To better understand the reason why nonlocal frequency shifts do not usually affect the width of the Rabi splitting, one can recall once more the CHO model of Eq.~(\ref{Eq:CHO}), according to which the energy difference between the two hybrid modes is
\begin{equation}\label{Eq:DiffE}
\Delta E = \sqrt{E_{\mathrm{R}}^{2} + \left(\hbar \omega_{\mathrm{p}} - \hbar \omega_{\mathrm{exc}} \right)^{2} }~,
\end{equation}
where $E_{\mathrm{R}} = \hbar \Omega_{\mathrm{R}}$. Nonlocal effects introduce a small correction to the plasmon, $\hbar \omega_{\mathrm{p}} \rightarrow \hbar \omega_{\mathrm{p}} + \delta E$. In the strong-coupling regime, where $\omega_{\mathrm{p}} \simeq \omega_{\mathrm{exc}}$, and assuming at a first step that $E_{\mathrm{R}}$ remains constant for simplicity, Eq.~(\ref{Eq:DiffE}) becomes
\begin{equation}\label{Eq:SmallDiffE}
\Delta E \simeq \sqrt{E_{\mathrm{R}}^{2} + \delta E^{2} } = E_{\mathrm{R}} \sqrt{1 + \left(\frac{\delta E}{E_{\mathrm{R}}}\right)^{2}}~.
\end{equation}
For Rabi splittings around 0.2\,eV, and nonlocal blueshifts of the order of 0.03\,eV, typical values in the examples explored here, this introduces a change in $\Delta E$ of about 1\%, which is hard to observe. According to the above analysis, one needs to identify systems with much larger nonlocal frequency shifts, accompanied with narrow Rabi splittings, to obtain a strong effect. What is still missing however is the influence of nonlocality on the coupling strength, and therefore, through $g = 2\hbar \Omega_{\mathrm{R}}$, on the Rabi splitting itself. In a semi-classical approach, and as long as the exact positions and orientations of individual emitters in the excitonic layer can be disregarded, the coupling strength is given by \cite{zengin_prl114}
\begin{equation}\label{Eq:CouplStrengh}
g = \mu \sqrt{\frac{N \hbar \omega}{2 \varepsilon \varepsilon_{0} V}}~,
\end{equation}
where $N$ is the number of emitters characterised by a transition dipole moment $\mu$, $\omega$ is the transition frequency, $\varepsilon_{0}$ the vacuum permittivity and $V$ the mode volume. For spherical particles, it has been shown analytically that, for the dipolar mode which is of interest here, the mode volume depends only on the the geometrical volume and the environment through $V = \left(4 \pi R^{3}/3\right) \left(1 + 1/2\varepsilon\right)$ \cite{khurgin_josab26}. Nevertheless, this result was obtained assuming local response functions for the metal, and nonlocal corrections to $V$ should be evaluated through the general definition \cite{maier_oex14}
\begin{equation}\label{Eq:GeneralV}
V = \frac{\int d^{3} \mathbf{r} \; u(\mathbf{r})}{\mathrm{max} \left\{ u(\mathbf{r}) \right\}}~,
\end{equation}
where $u (\mathbf{r})$ is the electromagnetic energy density at position $\mathbf{r}$. Introducing nonlocal corrections into Eq.~(\ref{Eq:GeneralV}) has been discussed in detail by Toscano \emph{et al.} \cite{toscano_nanophot2}. Our calculations have shown that within the GNOR model $V$ typically increases by no more than 20\%, which translates into a $\approx 10\%$ decrease in the coupling strength. Since this modification enters Eq.~(\ref{Eq:DiffE}) through a $E_{\mathrm{R}} \rightarrow E_{\mathrm{R}} - \delta E_{\mathrm{R}}$ correction, the nonlocal changes in coupling strength and plasmon energy tend, to first order, to cancel each other out. This tendency is further strengthened once absorptive losses are taken into account, which, according to Eq.~(\ref{Eq:CHOlossy}) introduce an extra damping correction $\delta \gamma$, working towards the same direction as $\delta E$ in Eq.~(\ref{Eq:SmallDiffE}), to further counterbalance the (relatively stronger) effect of increasing the mode volume. In view of the above discussion, interpreting the results of Figs.~\ref{fig1}--\ref{fig4} and understanding why nonlocal effects do not usually play an important role in the exciton-plasmon coupling can be achieved in a simple and intuitive manner.

\section{Conclusions}\label{Sec:conclusion}

In summary, we explored the influence of nonlocal effects in the description of the metal on the coupling of plasmonic nanoparticles with dye layers characterised by an excitonic transition. Through detailed simulations, in conjunction with analytical modelling, we showed that, contrary to expectations based mainly on results for single emitters in plasmonic environments, neither nonlocal frequency shifts due to screening, nor surface-enhanced Landau damping produces strong deviations from a description within classical electrodynamics. Apart from extreme situations of dramatic nonlocal blueshifts in situations of already narrow Rabi splittings, nonlocality affects plasmon-exciton coupling only incrementally, as long as the excitonic material is sufficiently described as a homogeneous layer. For simplicity, we have used here the GNOR model, leaving quantum corrections based on the Feibelman parameters for the centroid of charge or on surface-dipole moments for future investigation \cite{yan_prl115,christensen_prl118}. By analysing how nonlocal corrections in plasmonics enter into the common coupled-oscillator model, we provided a simple, intuitive interpretation of our findings. Novel plexcitonic architectures involving reduced, few-nm length scales, can be analysed without the necessity to resort to elaborate models going beyond classical electrodynamics, thus providing additional flexibility in the design and optimisation of systems with strong light-matter interactions.

\section{Methods}\label{Sec:methods}

Homogeneous silver nanoparticles and thin silver shells are described by the experimental dielectric function $\varepsilon_{\mathrm{exp}}$ of Johnson and Christy \cite{johnson_prb6}. When nonlocal corrections are introduced in the metal, the Drude part is subtracted from $\varepsilon_{\mathrm{exp}}$ to produce the background contribution to the metal dielectric function, $\varepsilon_{\infty}$, according to
\begin{equation}
\begin{split}
\varepsilon_{\mathrm{Drude}} &= \varepsilon_{\infty} - \frac{\omega_{\mathrm{p}}^{2}}{\omega \left(\omega + \mathrm{i} \gamma_{\mathrm{p}} \right)} \\
\varepsilon_{\infty} &= \varepsilon_{\mathrm{exp}} + \frac{\omega_{\mathrm{p}}^{2}}{\omega \left(\omega + \mathrm{i} \gamma_{\mathrm{p}} \right)}~,
\end{split}
\end{equation}
where for silver we use $\hbar \omega_{\mathrm{p}} =$ 8.99\,eV, $\hbar \gamma_{\mathrm{p}} =$ 0.025\,eV \cite{raza_jpcm27}.

In the case of single nanoparticles, we include nonlocal corrections in the scattering matrix of the Mie solution as described in Ref.~\cite{tserkezis_scirep6}. For a homogeneous metallic sphere of radius $R$ described by a dielectric function $\varepsilon_{1}$ in a homogeneous medium of $\varepsilon_{2}$ the electric-type nonlocal Mie coefficients of multipole order $\ell$, $t_{\ell}^{TM}$, are given by
\begin{equation}\label{Eq:NonlocalMie}
t_{\ell}^{TM} = \frac{-\varepsilon_{1} j_{\ell} (x_{1}) \left[x_{2} j_{\ell} (x_{2})\right]' + \varepsilon_{2} j_{\ell} (x_{2}) \left\{
\left[x_{1} j_{\ell} (x_{1}) \right]' + \Delta_{\ell} \right\}}
{\varepsilon_{1} j_{\ell} (x_{1}) [x_{2} h_{\ell}^{+} (x_{2})]' - \varepsilon_{2} h_{\ell}^{+} (x_{2}) \left\{ \left[x_{1} j_{\ell} (x_{1})\right]' + \Delta_{\ell} \right\} }~,
\end{equation}
where $j_{\ell} (x)$ and $h_{\ell}^{+} (x)$ are the spherical Bessel and first-type Hankel functions, respectively, while $x_{i} =
q_{i} R$ with $q_{i}$ being the (transverse) wave number in medium $i$ and $'$ denotes derivation with respect to the argument. The nonlocal correction $\Delta_{\ell}$ to the Mie coefficients is given as
\begin{equation}\label{eq:CNL}
\Delta_{\ell} = \ell \left(\ell + 1\right) j_{\ell} (x_{1})
\frac{\varepsilon_{1} - \varepsilon_{\infty}}{\varepsilon_{\infty}}
\frac{j_{\ell} (x_{\mathrm{L}})}{x_{\mathrm{L}}j_{\ell}'(x_{\mathrm{L}})} ~, 
\end{equation}
where $x_\mathrm{L} = q_{\mathrm{L}} R$ and $q_{\mathrm{L}}$ is the longitudinal wave number in the sphere, associated with the longitudinal dielectric function $\varepsilon_{\mathrm{L}}$, which is frequency- and wave vector-dependent \cite{ruppin_prl31,abajo_jpcc112}. The dispersion of longitudinal waves is given by $\varepsilon_{\mathrm{L}} (\omega, \mathbf{q}) = 0$. In the limiting case where $\Delta_{\ell} = 0$ we retrieve the local result of standard Mie theory. The corresponding analytical solution is lengthy and can be found in the Supplementary Information of Ref.~\cite{tserkezis_nscale8}.

In the case of dimers, we use a commercial finite-element solver (Comsol Multiphysics 5.0) \cite{comsol}, appropriately adapted to include the description of nonlocal effects \cite{nanoplorg}, to solve the system of coupled equations \cite{mortensen_natcom5}
\begin{equation}
\begin{split}
& \nabla \times \nabla \times \mathbf{E} (\mathbf{r}, \omega) = \left(\frac{\omega}{c}\right)^{2} \varepsilon_{\infty} \mathbf{E} (\mathbf{r}, \omega) + \mathrm{i} \omega \mu_{0} \mathbf{J} (\mathbf{r}, \omega)\\
& \left[\frac{\beta^{2}}{\omega \left(\omega + \mathrm{i} \gamma\right)} + \frac{D}{\mathrm{i} \omega} \right] \nabla \left[\nabla \cdot \mathbf{J} (\mathrm{r}, \omega) \right] + \mathbf{J} (\mathrm{r}, \omega) = \sigma \mathbf{E} (\mathrm{r}, \omega)~,
\end{split}
\end{equation}
where $\mathbf{E}$ and $\mathbf{J}$ are the electric field and the induced current density respectively; $\sigma = \mathrm{i} \varepsilon_{0} \omega_{\mathrm{p}}^{2}/(\omega + \mathrm{i} \gamma_{\mathrm{p}})$ is the Drude conductivity, and $\mu_{0}$ is the vacuum permeability, related to the velocity of light in vacuum through $c = 1/\sqrt{\varepsilon_{0} \mu_{0}}$.

In both approaches, the hydrodynamic parameter $\beta$ is taken equal to $\sqrt{3/5} \; v_{\mathrm{F}}$, where $v_{\mathrm{F}} = 1.39 \cdot 10^{6}$ m s$^{-1}$ is the Fermi velocity of silver \cite{raza_jpcm27}, while for the diffusion constant $D$ we use $D = 2.684 \cdot 10^{4}$ m$^{2}$ s$^{-1}$ \cite{tserkezis_scirep6}. Values of this order of magnitude reproduce well the experimentally observed size-dependent broadening in small metallic nanoparticles \cite{kreibig_surfsci156}. As the additional boundary condition we adopt the usual condition of zero normal component of the induced current density at the metal boundary \cite{raza_jpcm27}, which implies a hard-wall description of the metal, an approach which might prove inefficient in the case of good jellium metals \cite{teperik_prl110,ciraci_prb95}, but is reasonable for a noble metal with high work function like silver. When describing the interface between silver and the excitonic layer, we assume that carriers from the one medium are not allowed to enter the other, an assumption justified by the fact that the excitonic layer consists of an assembly of dye molecules, and is only effectively described as a homogeneous layer.

For the excitonic material we use a Drude--Lorentz model according to
\begin{equation}\label{Eq:DrudeLorentz}
\varepsilon = 1 - \frac{f \omega_{\mathrm{exc}}^{2}}{\omega^{2} - \omega_{\mathrm{exc}}^{2} + \mathrm{i}\omega \gamma_{\mathrm{exc}}}~,
\end{equation}
where $f$ is the reduced oscillator strength. Throughout the paper $\omega_{\mathrm{exc}}$ is allowed to vary as stated in each case, $f =$ 0.02, and $\hbar \gamma_{\mathrm{exc}} =$ 0.052\,eV \cite{fofang_nl8}.

\begin{acknowledgements}
C.~T. was supported by funding from the People Programme (Marie Curie Actions) of the European Union's Seventh Framework Programme (FP7/2007-2013) under REA grant agreement number 609405 (COFUNDPostdocDTU).
We gratefully acknowledge support from VILLUM Fonden via the VKR Centre of Excellence NATEC-II and from the Danish Council for Independent Research (FNU 1323-00087). The Center for Nanostructured Graphene (CNG) was financed by the Danish National Research Council (DNRF103).
N.A.M. is a VILLUM Investigator supported by  VILLUM Fonden.
\end{acknowledgements}

\end{document}